\documentclass[aps,twocolumn,showpacs,floatfix]{revtex4}
\usepackage{amsmath,amsfonts,amssymb,graphicx,color,times,bbm,psfrag,dsfont}
\usepackage{epsfig}
\usepackage[T1]{fontenc}

\newcommand{\be}{\begin{equation}}
\newcommand{\ee}{\end{equation}}
\newcommand{\bra}[1] {\langle{#1}\vert }
\newcommand{\ket}[1] {\vert{#1}\rangle }

\def\trace{{\rm tr}\;}
\def\nbOne{{\mathchoice {\rm 1\mskip-4mu l} {\rm 1\mskip-4mu l}{\rm 1\mskip-4.5mu l} {\rm 1\mskip-5mu l}}}

\begin{document}

\title{A short review on entanglement in quantum spin systems}

\author{J. I. Latorre and A. Riera}

\affiliation{
Dept. d'Estructura i Constituents de la Mat\`eria,
Universitat de Barcelona, 647 Diagonal, 08028 Barcelona, Spain
}

\begin{abstract}
We review some of the recent progress on the study of entropy of
entanglement in many-body quantum systems.
Emphasis is placed on the scaling properties of entropy for
one-dimensional multi-partite models at
quantum phase transitions and, more generally, on the concept of area law.
We also briefly describe the relation between entanglement
and the presence of impurities, the idea of particle entanglement,
the evolution of entanglement along renormalization
group trajectories, the dynamical evolution of entanglement and the
fate of entanglement along a quantum computation.
\end{abstract}
\maketitle

\tableofcontents

\section{Introduction}
Quantum systems are ultimately characterized by the observable correlations they exhibit. 
For instance, an observable
such as the correlation function between two spins in a typical spin chain 
may decay exponentially as a function of
the distance separating them or, in the case the system undergoes a phase transition, 
algebraically. The correct assessment of these quantum correlations is tantamount to understanding 
how entanglement is distributed in the state of the system. This is easily understood as follows. Let us 
consider a connected correlation 
\begin{align}
\langle \Psi|& O_i O_j|\Psi\rangle_c\equiv \\ \nonumber
&\langle \Psi| O_i O_j|\Psi\rangle - \langle \Psi| O_i|\Psi\rangle \langle \Psi| O_j|\Psi\rangle\ ,
\end{align}
where $O_i$ and $O_j$ are operators at sites $i$ and $j$ respectively.
This connected correlator would vanish identically for any product state $|\Psi\rangle=
\otimes_i |\psi_i\rangle$. That is, $O_i\otimes O_j$ is a product operator
and, consequently, its correlations can only come from the amount
of entanglement in the state $|\Psi\rangle$. It follows that the
ground state of any interesting system will be highly correlated 
and, as a particular case, even the vacuum displays a non-trivial entanglement
structure in quantum field theories.

Notice that, at this point, our emphasis
has moved from Hamiltonians to states. It is perfectly sensible to analyse
entanglement properties of specific states per se, which 
may be artificially created using a post-selection mechanism or may effectively
be obtained in different ways using various interactions. We are, thus, concerned
with the entanglement properties that characterize a quantum state.  Yet, 
we shall focus on states that are physically relevant. In particular, we
shall study the entanglement properties of ground states of Hamiltonians
that describe the interaction present in spin chains.

It is clear that the property of entanglement can be made apparent by
studying correlations functions on a given state. We could consider two-, three- or n-point connected
correlation functions. Any of them would manifest how the original interactions in the
Hamiltonian have operated in the system to achieve the observed degree of entanglement.
For instance, free particles (Gaussian Hamiltonians) produce n-point correlators
that reduce to products of two-point correlators {\sl via} Wick's theorem.
Nonetheless, the study of specific correlation functions is model dependent. How can
we compare the correlations of a Heisenberg Model with those of Quantum Chromodynamics?
Each theory brings its own set of local and non-local operators that close an Operator Product Expansion.
Different theories will carry different sets of operators, so that a naive
comparison is hopeless. A wonderful possibility to quantify
degrees of entanglement for unrelated theories emerges from the use of Renormalization Group ideas and
the study of universal properties. For instance, a system may display exponential
decays in its correlations functions which is globally controlled by a common correlation length. 
A model with a larger correlation
length is expected to present stronger long-distance quantum correlations. 

We may as well try to find a
universal unique figure of merit that would allow 
for a fair comparison of the entanglement present in {\sl e. g.} the
ground state of all possible theories. 
Such a figure of merit
cannot be attached to the correlations properties of model-dependent operators
since it would not allow for comparison among different theories.
The way to overcome this problem is to look for an operator which is defined
in every theory. It turns out that there is only one such operator: the stress
tensor. To be more precise, we can use the language of conformal field
theory which establishes that there always is a highest weight operator
that we call the Identity. The Identity will bring a tower of descendants,
the stress tensor being its first representative. Indeed, the stress tensor
is always defined in any theory since it corresponds to the operator
that measures the energy content of the system and it is the operator
that couples the system to gravity. Correlators of stress tensor operators
are naturally related to entanglement. In particular, the coefficient
of the two-point stress tensor correlator in a conformal field theory
in two dimensions corresponds to the central charge of the theory.

There is a second option to measure entanglement in a given state
with a single measure of entanglement which is closer in spirit
to the ideas of Quantum Information. The basic idea consists
of using the von Neumann (entanglement) entropy of the reduced density matrix
of a sub-part of the system which is analysed. Indeed,
the entanglement entropy quantifies the amount of surprise that a
sub-part of a system finds when discovering it is correlated to the
rest of the system. Therefore, entanglement entropy is a {\sl bona fide}
measure of the correlations in the system. The advantage of 
the von Neumann entropy of entanglement is that it can be defined
for any system. We expect its general properties, as the
way it scales with the size of the sub-part of the system we are considering,
should characterized the quantum state in a quite refined way.
 
It is tantalizing to exhaustively explore the behaviour of
the entropy of entanglement in relevant physical systems. For instance,
will the entropy of entanglement scale differently at 
a critical point as compare to a non-critical one? Will
scaling properties depend on the dimensionality of the system.
Is disorder relevant for long-distance correlations?
Are there non-local systems where entropy obeys some singular
behaviour? How does entanglement renormalize?
How does entanglement evolve dynamically?  We can even
go further away from standard dynamical models and question
whether entanglement is somehow related to computational
complexity problems, both NP-complete and QMA-complete. 
We shall now briefly review some of these questions.

\section{An explicit computation of entanglement entropy}
Let us start our discussion with the study of the behaviour of entanglement
at different regimes (critical and non-critical) of the XX model.
As we shall see, 
entanglement entropy will be a good tool to describe
the properties of the quantum phase transition which characterize this model
\cite{vid02,LRV04}.

In order to do this, first we need to introduce the Von Neumann entropy as a
measure of the bipartite entanglement in pure sates.
Then, we will study the scaling of the entanglement entropy for the simple XX model. 
We then will proceed to compute the ground state $\ket{GS}$ of the system, 
from which we can obtain the spectra of the reduced density matrix $\rho_L$ 
for the block of $L$ contiguous spins. 
The knowledge of the eigenvalues of $\rho_L$ will let us determine
its entanglement entropy $S_L$.
Finally, we are going to analyse 
how the entanglement behaves depending on the critical properties of
the model.

\subsection{Entanglement Entropy}

The problem of measuring and quantifying quantum correlations, or {\sl entanglement}, 
in many body quantum systems is a field of research in its own, that
benefits both from condensed matter and quantum information ideas. 
Here, we shall only discuss the Von Neumann entropy as a figure of merit 
for entanglement. Nevertheless,
there are  many other measures that have been largely explored.
A detailed explanation of them can be found in
several reviews 
\cite{AFOV08, PV07, Horodecki07, Eisert06, BZ06, Vedral02, Bruss02, Wootters01, PV98}.
Our choice for the entropy of entanglement is based on a combination
of ideas. Entropy has a clear information theory meaning.
It also relates to extensive research in quantum field theory and
the physics of black holes. Furthermore, its scaling properties are 
related to the characterization
of quantum phase transitions as provided by conformal field theory.  
On the other hand, entanglement entropy is not a simple quantity to
compute, neither a direct observable (though it relates to them).

The von Neumann entropy can be used to measure the entanglement between two parts of the
system, that we call $A$ and $B$. Let us take a ket
$\ket{\psi}_{AB}$ belonging to $\cal{H}=\cal{H}_A\otimes \cal{H}_B$.
According to the Schmidt decomposition, for any pure bipartite state 
we can always find
two orthonormal basis $\{ \ket{\varphi_i}_A \}$ and $\{ \ket{\phi_j}_B\}$ 
such that the state $\ket{\psi}_{AB}$ can be written as
\be
\ket{\psi_{AB}}=\sum_{i}^\chi\alpha_{i}\ket{\varphi_i}_A\ket{\phi_i}_B \, ,
\label{Schmidt}
\ee
where  $\alpha_i$ can be chosen real and positive and are called Schmidt coefficients,
and $\chi\le\min\left({\rm dim} {\cal H}_A, {\rm dim} {\cal H}_B\right)$ is the
Schmidt number. Note that the Schmidt decomposition is just
the diagonalization of the matrix of coefficients in the original state
which is always possible if we can perform two independent unitary
transformations in $A$ and $B$.

The Von Neumann entropy between these bipartitions is defined as 
the Shannon entropy of the square of the Schmidt coefficients,
\begin{equation}
S_A= S_B\equiv -\sum_i \alpha_i^2 \log\alpha_i^2 \; .
\end{equation}

This expression can be written in terms of the reduced density matrices
of each part of the system. That is, 
\be
S_A\equiv S(\rho_A)=-\trace\left( \rho_{A} \log_2 \rho_{A} \right) \, ,
\label{Eq:entropy-def}
\ee
where 
\be
\rho_{A} = \trace_{B}(\ket{\psi}_{AB}\bra{\psi}_{AB})=\sum_{i} \alpha_{i}^2\ket{\phi_i}_B
\bra{\psi_{i}}_B \, . 
\ee
It is easy to see that $S_A=S_B$. Thus, the surprise that A experiences
when discovering its correlation to B is identical to the one of B
realizing its correlation with A.

The von Neumann entropy  verifies the following properties:
{\it i})  it is invariant under local unitary operations 
($S_A=S_B$ is  a function of the $\alpha_i$'s only);
{\it ii})  it is continuous (in a certain sense also in the asymptotic 
    limit of infinite copies of the state; see e.g. Ref.\ \cite{PV07});
{\it iii})  it is additive:
    $S(\ket{\psi}\otimes \ket{\phi}) = 
    S(\ket{\psi})+S(\ket{\phi})$.

In our particular case, we are going to use the entanglement entropy to 
study the quantum correlations of spin chains.
We will be interested in determine the entanglement between a block
of $L$ contiguous spins and the rest of the chain. 
Then, if $\ket{GS}$ represents the ground state of a system of $N$ spins, 
$\rho_{L} = \trace_{N-L}(\ket{GS}\bra{GS})$ is the reduced density matrix
of the block of $L$ contiguous spins that we will use in Eq.\ \ref{Eq:entropy-def}.

Finally, let us point out that, in the case the ground state 
that we study is translationally invariant, neither $\rho_L$
nor $S_L$ will depend on the position of the block of spins in the chain.
In this case, it is easy to show that the entropy $S_L$ 
is a concave function respect to $L$ \cite{OP93}
\be 
S_L \geq \frac{S_{L-M}+S_{L+M}}{2},
\label{Eq:concavity}
\ee 
where $L=0, \cdots, N,$ and $M=0, \cdots,\min~\{N\!-\!L,L\}$.

\subsection{XX model}

We shall now present a computation of entanglement entropy for the
reduced density matrix of the ground state of the widely studied XX model \cite{vid02,LRV04}.
This theory captures the non-trivial structure of a quantum
phase transition, while remaining simple enough to carry
explicit computations  throughout.
 The XX model consists of a chain of $N$ spin $\frac{1}{2}$ particles with nearest 
neighbour interactions and an external magnetic field. 
Its Hamiltonian is given by
\be
H_{XX} = - \frac{1}{2}\sum_{l=0}^{N-1} \left(  \sigma_l^x \sigma_{l+1}^x +
  \sigma_l^y \sigma_{l+1}^y \right) + \frac{1}{2}\lambda \sum_{l=0}^{N-1}\sigma_l^z   \, ,
\label{eq:XX-Hamiltonian}
\ee
where $l$ labels the $N$ spins, $\lambda$ is the magnetic field and $\sigma_l^{\mu}$ $(\mu=x,y,z)$ 
are the Pauli matrices at site $l$.

Without loss of generality, we are going to consider that the magnetic field 
is oriented in the positive Z direction ($\lambda>0$), since, if this was not
the case, we could always map the system onto an equivalent one with $\lambda>0$ 
by simply interchanging the spin states up and down.

\subsection{Ground State}
Next, we need to compute the ground state $\ket{GS}$ of the XX Hamiltonian (\ref{eq:XX-Hamiltonian}).
In order to do this, we will follow two steps: 
({\it i}) first, we will perform a Jordan-Wigner transformation
to  rewrite $H_{XX}$ as a quadratic form of fermionic operators,
and then ({\it ii}) we will take profit of the translational invariance of the system 
realizing a Fourier transform which will diagonalize the Hamiltonian.
A third step which is needed in the more general
XY model, the Bogoliubov transformation, is not necessary
in this particular case.
Let us remark that this computation is standard and 
appears in many text books \cite{sach99,hen01,chak01}.

The Jordan-Wigner transformation maps a spin chain 
of interacting qubits onto an equivalent system of interacting fermions.
This powerful transform is defined by the following relation between the Pauli matrices
and the creation and annihilation of the fermionic modes
\be
a_l = \left( \prod_{m=0}^{l-1}\sigma_m^z \right) \frac{\sigma_l^x - i\sigma_l^y}{2} \, .
\label{Eq:JW-definition}
\ee
We, indeed, can check that the fermionic operators $a_l$ fulfil 
the canonical commutation relations
\be
\{a_l^{\dagger},a_m\}=\delta_{lm} , ~~~\{a_l,a_m\} =0 \, .
\ee
The idea behind the transformation is to identify the state of the spin $l$ (0 or 1 in the computational basis)
with the occupation number of the corresponding fermionic mode. 
Thus, in Eq.\ \ref{Eq:JW-definition}, the factor $(\sigma_l^x - i\sigma_l^y)/2$ corresponds to
the operator $\ket{0}\bra{1}$ in the computational basis, 
and the product $\prod_{m=0}^{l-1}\sigma_m^z$  generates the appropriate sign
in order to satisfy the commutation relations.

The Jordan-Wigner transformation casts the XX Hamiltonian onto
\be
H_{XX}= -\sum_{l=0}^{N-1} \left( a_l^\dagger a_{l+1} + a_{l+1}^\dagger a_l  \right)
+\lambda\sum_{l=0}^{N-1}  a_l^\dagger a_l \, ,
\ee
which corresponds to a model of free fermions with chemical potential $\lambda$.

Now, let us exploit the translational symmetry of the system by introducing
the Fourier transformed fermionic operators 
\be
b_k =\frac{1}{\sqrt{N}} \sum_{l=0}^{N-1} a_l e^{-i\frac{2\pi}{N}kl},
\ee 
where $0\le k \le N-1$. As the Fourier transform is a unitary transformation, 
these new $b_k$ operators also satisfy the canonical commutation relations and, therefore, 
they are fermionic operators.

The Hamiltonian, written in terms of these $b_k$ operators, displays a diagonal structure
\be
H_{XX}=\sum_{k=0}^{N-1} \Lambda_k b_k^\dagger b_k \, ,
\ee
where the energy that penalizes (or favours, depending on the sign) the
occupation of mode $k$ is
\be
 \Lambda_k=\lambda-2 \cos \frac{2\pi k}{N} \, .
\label{Eq:Lambda_k}
\ee
We have assumed that the system satisfied periodic boundary conditions.
If this was not the case, the Hamiltonian would not be diagonal due to an extra term 
proportional to $\frac{1}{N}$. 
In the thermodynamic limit, therefore, this extra term disappears.

We realize that, on one hand, if $\lambda > 2$, then $\Lambda_k \ge 0$ $\forall \ k$. 
This implies that the ground state of
the system is the state annihilated by all $b_k$ operators
\be
b_k \ket{GS} = 0 \ \ \forall \ k \, ,
\label{Eq:gs-A}
\ee
and, therefore, it has 0 energy. 

\begin{figure}[!ht]
\scalebox{0.6}{\includegraphics{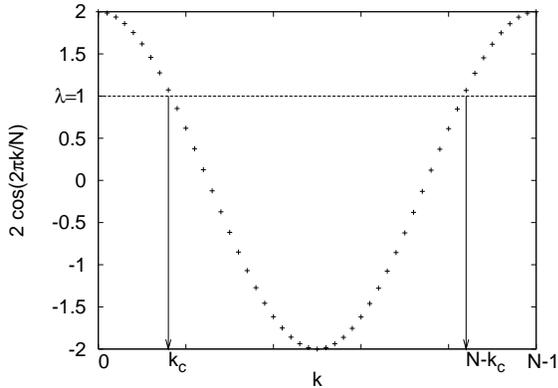}}
\caption{The two terms of $\Lambda_k$, Eq.\ (\ref{Eq:Lambda_k}), are plotted
for the particular case $\lambda=1$.
We realize that if $2\cos \left(\frac{2\pi k}{N}\right)>\lambda$, $\Lambda_k <0$ 
while if $2\cos \left(\frac{2\pi k}{N}\right)<\lambda$, $\Lambda_k >0$.} 
\label{fig:ground-state}
\end{figure}

On the other hand, if $2 > \lambda \ge 0$, the ground state is the state 
annihilated by the operators $b_k$ with $\Lambda_k > 0$ and $b_m^\dagger$
with $\Lambda_m<0$,
\begin{align}
b_k \ket{GS} &= 0 \ \ \textrm{if} \ \ \Lambda_k>0 \nonumber \\
b^\dagger_m \ket{GS} &= 0 \ \ \textrm{if} \ \ \Lambda_m<0  \, ,
\label{Eq:gs-B}
\end{align}
and its energy is simply $\sum_m \Lambda_m$ $\forall \ \Lambda_m<0$.
In Fig.\ \ref{fig:ground-state} and Eq.\ (\ref{Eq:Lambda_k}), we can see that 
if $k_c\ge k \ge 0$ or $N-1\ge k \ge N-k_c$, where $k_c$ is defined by
\be
k_c=\left[ \frac{N}{2\pi}\arccos \left( \frac{\lambda}{2}\right)\right] \, ,
\label{Eq:k_c}
\ee
then  $\Lambda_k<0$,
whereas for the rest of cases $\Lambda_k\ge 0$. In Eq.\ (\ref{Eq:k_c}), the
brackets [] represent the {\sl floor} function.

\subsection{Entanglement entropy of a block}
The strategy to get the Von Neumann entropy of a block of $L$ spins first consists
in computing the correlation matrix $\langle a_m^\dagger a_n \rangle$ of the GS in this block. 
Then, the eigenvalues of this correlation matrix are related  with the eigenvalues of 
the reduced density matrix of the block which are 
required to determine the entanglement entropy.

The simple structure of the GS, shown in Eqs.\ (\ref{Eq:gs-A}) and (\ref{Eq:gs-B}),
makes easy to compute its correlation matrix 
\be
\langle b_p^\dagger b_q \rangle = \left\{ \begin{array}{ll}
\delta_{pq} & \textrm{if } \Lambda_p<0 \\
0 & \textrm{if } \Lambda_p>0
\end{array}\right.
\, .
\ee
From now on, we will consider the case in which $2>\lambda \ge 0$.
Notice that if $\lambda >2$, then $\langle b_p^\dagger b_q \rangle=0$ for all $p$ and $q$.
This case is trivial to analyse, since the correlators $\langle a_m^\dagger a_n \rangle$ are also
null, and the GS is in a product state. 

The next step is  to go back in the Fourier transform to get the correlation matrix of the $a_n$ operators
\be
\langle a_m^\dagger a_n \rangle = \frac{2}{N} \sum_{k=0}^{k_c} \cos \left[ \frac{2 \pi }{N} k (m-n)\right]
\, .
\ee
In the thermodynamic limit, the previous sum becomes an integral and it can 
be determined analytically. 
In this case, the correlation matrix of the block of $L$ fermions in position space is, 
\be
A_{mn}= \langle a_m^\dagger a_n \rangle= \frac{1}{\pi}\frac{\sin k_c(m-n)}{m-n} \, ,
\label{Eq:correlation-matrix-block}
\ee
where $L\ge m,n \ge 0$.
Notice that, by means of Wick's theorem,
any operator that acts on the block 
can be written in terms of the correlation matrix $A_{mn}$.
For instance,
\be
\langle a_k^\dagger a_l^\dagger a_m a_n\rangle = 
\langle a_k^\dagger a_n\rangle \langle a_l^\dagger a_m\rangle
-\langle a_k^\dagger a_m\rangle \langle a_l^\dagger a_n\rangle \, .
\ee
This is due to the fact that the system is Gaussian, and its eigenstates are 
determined by the first and second moments of some fermionic operators.

The correlation matrix $A_{mn}$ could also be computed using the density matrix of the block 
$\rho_L$, 
\be
A_{mn}=\textrm{Tr}(a_m a_n \rho_L) \, .
\ee
We, thus, need to invert the previous equation, that is, to compute the density matrix
$\rho_L$ from the correlation matrix $A_{mn}$.

The matrix $A_{mn}$ is Hermitian, so it can be diagonalized by a unitary transformation,
\be
G_{pq}=\sum_{m,n=0}u_{pm}A_{mn}u^*_{nq}=\langle g^\dagger_p g_p\rangle \delta_{pq} \, ,
\ee
where $g_p=\sum_m u_{pm} a_m$.
In this case, the density matrix of the block must also verify
\be
G_{mn}=\textrm{Tr}(g^\dagger_m g_n \rho_L)= \nu_m \delta_{mn} \, ,
\label{Eq:corr-diag}
\ee
which implies that $\rho_L$ is uncorrelated and it can be written as
\be
\rho_L=\varrho_1 \otimes \cdots \otimes \varrho_L \, ,
\ee
where $\varrho_m$ is the density matrix corresponding to the $m$-th 
fermionic mode excited by $g^\dagger_m$. 

In order to determine the eigenvalues of the density matrix of one mode,
let us express the $g_m$, $g^\dagger_m$ and $\varrho_m$ operators in its matrix representation.
That is,
\be
g_m=
\left(\begin{array}{cc}
0 & 0 \\ 1 & 0
\end{array} \right)
\ \ \ \ 
g_m^\dagger=
\left(\begin{array}{cc}
0 & 1 \\ 0 & 0
\end{array} \right) \, ,
\ee
and 
\be
\varrho_m=
\left(\begin{array}{cc}
\alpha_m & \beta_m \\ \beta_m^* & 1-\alpha_m 
\end{array} \right) \, ,
\ee
where $\alpha_m$ and $\beta_m$ are the matrix elements of $\varrho_m$
that we want to determine.
It is easy to see that $\beta_m=0$, since
\be
\langle g_m \rangle = \textrm{Tr}(g_m \rho_L)= \beta_m = 0 \, .
\ee
Moreover, rewriting Eq.\ \ref{Eq:corr-diag} in terms of these matrices  
\be
\textrm{Tr}(g^\dagger_m g_m \rho_L)= 
\textrm{Tr}\left[
\left(\begin{array}{cc}
1 & 0 \\ 0 & 0
\end{array} \right)
\left(\begin{array}{cc}
\alpha_m & 0 \\ 0 & 1-\alpha_m 
\end{array} \right)\right]
=\nu_m
\ee
we realize that $\alpha_m=\nu_m$.

The entanglement entropy between the block and the rest of the system is
therefore,
\be
S_L = \sum_{l=1}^L H_2\left(\nu_l\right).
\label{Eq:entanglement-entropy-XX}
\ee
where $H_2(x)= -x\log x -(1-x)\log(1-x)$ denotes the binary entropy.

Summing up, the three steps that we have to follow in order to compute the
entanglement entropy of the GS of a block of $L$ spins for the XX model are:
({\it i}) to determine the correlation matrix $A_{mn}$ by evaluating 
Eq.\ (\ref{Eq:correlation-matrix-block}) for $L\ge m,n \ge 0$,
({\it ii}) to diagonalize this correlation matrix and, with its eigenvalues,
({\it iii}) to compute the entanglement entropy according to Eq.\ (\ref{Eq:entanglement-entropy-XX}).

Let us emphasize that this method is computationally efficient, since its computational cost
scales polynomially with the number of spins of the block $O(L^3)$, whereas 
the Hilbert space of the problem has dimension $2^N$.

It is also necessary to recall a quite subtle point that we have
skipped along our previous discussion.  It turns out that there is no need to perform a final
transformation back to spins, that is, there is no need to invert
the Jordan-Wigner transformation. This is due to the fact that the coefficients
of a given state are identical when written in terms of the spin basis  or
in terms of the fermionic $a_l$ operators. More precisely,
\begin{align}
\label{identicalcoefficients}
\vert\psi\rangle &=\sum_{i_1,i_2,\ldots,i_n}C^{ i_1,i_2,\ldots,i_n} \vert i_1,i_2,\ldots,i_n\rangle 
\\
&=\sum_{i_1,i_2,\ldots,i_n}C^{ i_1,i_2,\ldots,i_n}(a^\dagger_1)^{i_1}(a^\dagger_2)^{i_2}\ldots (a^\dagger_n)^{i_n} \vert \textrm{vac}\rangle \ . \nonumber
\end{align}
Thus, the same coefficients appear in the ket, either when written in
the initial spin basis, or when expressed as creation operators in the fermionic vacuum, $\vert \textrm{vac}\rangle$.
Consequently, the reduced density matrix entropy of entanglement is identical
for both expressions.

Finally, let us mention that the computation of the geometric entropy of
Gaussian systems have been systematized in Refs.\ \cite{Peschel}.
In particular, it is shown that for solvable fermionic and bosonic lattice systems, 
the reduced density matrices can be determined from the properties of the correlation
functions. This subject is reviewed in Ref.\ \cite{Peschel-review} in this
special issue.

\subsection{Scaling of the entropy}
It is now easy to compute the entanglement entropy of the ground state of the XX model
for arbitrary values of the block size $L$ and magnetic field $\lambda$.

\begin{figure}[!ht]
\scalebox{0.6}{\includegraphics{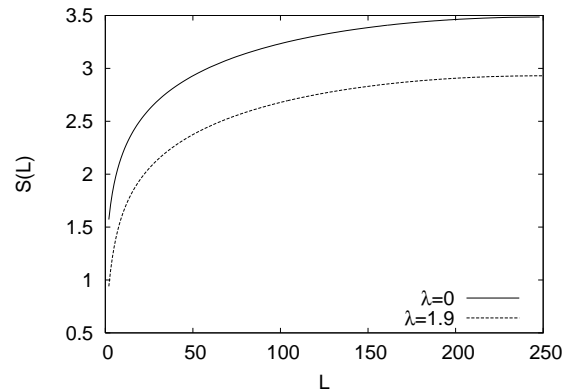}}
\caption{Entropy of the reduced density matrix of $L$
spins for the XX model in the limit $N\to\infty$, for two different values of the external
magnetic field $\lambda$. The maximum entropy is reached when there is
no applied external field ($\lambda=0$). The entropy decreases while the magnetic
field increases until $\lambda=2$ when the system reaches the
ferromagnetic limit and the ground state is a product state in the
spin basis.}
\label{fig:scaling-entropy-XX}
\end{figure}

In Fig.\ \ref{fig:scaling-entropy-XX}, we show how the entropy of the reduced density matrix 
of a block of $L$ spins varies with $L$ for different values of the magnetic field.
The maximum entropy is reached for $\lambda=0$. 
In particular, we recover the result in Ref.\ \cite{LRV04}
and see that for $\lambda=0$ the leading scaling of the entropy is perfectly fitted
by a logarithm,
\be
S_L=\frac{1}{3}\log_2 L + a \, , 
\label{eq:scaling-entropy-XX}
\ee
where $a$ is a constant that was determined analytically in Ref.
 \cite{IJK04}.

As we increase the magnetic field, but it is still less than 2, 
the entropy decreases although it keeps its logarithmic behaviour with $L$.  
When $\lambda>2$, the entropy saturates to zero, 
since the ground state is already in a product state
in the spin basis corresponding to the ferromagnetic phase, 
$\prod_i \ket{\uparrow}_i$.

The relation between logarithmic scaling and entropy
is confirmed by similar computations in different models.
The general result is that entanglement entropy obeys
a logarithmic scaling relation at critical points, 
that is when the system is at a phase transition, whereas
a saturation of entanglement is found away from criticality.
This universal logarithmic at critical points must
emerge from the basic symmetry that characterizes phase
transitions, namely conformal invariance. We shall come
to this developments in the next section.

Let us mention that the scaling of entanglement entropy was formerly studied
in the context of quantum field theory and black-hole physics. There, the system
sits in higher dimensions. The entanglement entropy
scales following an area law that we shall discuss later on.
Yet, it is important to note that one-dimensional systems
are an exception to the area law. Entanglement pervades
the system at any distance, not staying just at the point-like
borders of a block.

Summing up, we have seen that the scaling entropy is
a good witness for quantum phase transitions. 
Many other studies of different measures of entanglement 
at quantum phase transitions have been presented recently.
Let us here mention that in  Refs.\ \cite{Zanardi06, Zhou08}, 
quantum phase transitions are characterized in terms of the overlap (fidelity)
function between two ground states obtained for two close values of external parameters.
When crossing the critical point a peak of the fidelity is observed.

\subsection{Entanglement entropy and Toeplitz determinant}
Before finishing this section, we would like to sketch
how the particular structure of the correlation matrix in Eq.\ (\ref{Eq:correlation-matrix-block}) 
allows us to derive an analytical expression for the scaling 
law of the entanglement entropy. This result is presented in Ref.\ \cite{IJK04}.

In order to obtain an analytical expression for
the entanglement entropy, let us introduce the function
\be
D_{L}(\mu)= \det \left(\tilde{A}(\mu) \right) \, ,
\ee
where $\tilde{A}(\mu) \equiv \mu I_{L}- A$, $I_L$ is the identity matrix
of dimension $L$, and $A$ is the correlation matrix 
defined in Eq.\ (\ref{Eq:correlation-matrix-block}). 
If we express the matrix $A$ in its diagonal form, we trivially have
\be
D_{L}(\mu)=\prod_{m=1}^{L} (\mu-\nu_m),
\ee
where $\nu_m$, with $m=1,\ldots,L$, are the eigenvalues of $A$.
Then, according to Cauchy residue theorem, 
the entanglement entropy $S_L$ can be expressed in terms of an 
integral in the $\mu$-complex plane as follows 
\begin{align}
S_L&=\lim_{\epsilon \to 0^+} \lim_{\delta \to
0^+}\frac{1}{2\pi \mathrm{i}} \oint_{c(\epsilon,\delta)}
e(1+\epsilon, \mu) \mathrm{d} \ln D_{L}(\mu) \nonumber \\
&=\sum_{m=1}^L H_2(\nu_m)
\; ,\label{Eq:complex-integral}
\end{align}
where $c(\epsilon,\delta)$ is a closed path that encircles all zeros of $D_L(\mu)$ 
and $e(1+\epsilon, \mu)$ is an arbitrary function that is analytic in the contour $c(\epsilon,\delta)$
and verifies $e(1, \nu_m)=H_2(\nu_m) \ \ \forall \ m$. 

Thus, if we could obtain an analytical expression for the $D_L(\mu)$ function, 
we would be able to get a closed analytical result for the entanglement entropy. 

Notice that both the correlation matrix $A_{mn}$, defined in Eq.\ (\ref{Eq:correlation-matrix-block}),
and $\tilde A$ are Toeplitz matrices, 
that is to say, matrices in which each descending diagonal 
from left to right is constant,
\be
A=
\left( \begin{array}{cccc}
f_0 &f_{-1}& \ldots &f_{1-L}\\
f_{1}& f_0&   &   \vdots\\
\vdots &      & \ddots&\vdots\\
f_{L-1}& \ldots& \ldots& f_0
\end{array}     \right) \; ,
\ee
where, in this case, $f_m=\frac{1}{\pi}\frac{\sin k_c m}{m}$.

The asymptotic behaviour (when $L \to \infty$) of the determinant of Toeplitz matrices
has been widely studied in many cases, giving the famous Fisher-Hartwig conjecture
(see Refs.\ \cite{wu, fisher, basor, tracy, abbs}).
In our particular case, the expression for the determinant of $\tilde{A}$ 
was proven in Ref.\ \cite{basor} and, therefore, it is a theorem instead of a conjecture.
In this way, we may insert this result in Eq.\ (\ref{Eq:complex-integral}), perform
the corresponding complex integral and obtain the asymptotic analytical expression for the
entanglement entropy of the XX model. This computation is presented explicitly in 
Ref.\ \cite{IJK04} with the final result,
\be
S_L=\frac{1}{3} \ln \mathrm{L} +\frac{1}{6} \ln
\left(1-\left(\frac{\lambda}{2}\right)^2\right) +\frac{\ln 2}{3}
+\Upsilon_1,
\ee
where
\begin{equation}
  \Upsilon_1 =   - \int_0^\infty \mathrm{d} t \left\{ {e^{-t} \over 3 t}
  + {1 \over  t \sinh^2 (t/2)} - { \cosh (t/2) \over 2 \sinh^3 (t/2)}
  \right\}. 
\end{equation}
Indeed, we realize that this analytical expression for the scaling of the entanglement
entropy is compatible with the numerical fit
of Eq.\ (\ref{eq:scaling-entropy-XX}) and, moreover, 
it fixes the value of the independent term.

This procedure is also used to obtain an analytical expression for the entanglement entropy
of the XY model in Refs. \cite{Kor}.
In Ref. \cite{Kor-RenyiEntropy}, the scaling of the Renyi entropy is determined for the XY model in terms of Klein's elliptic $\lambda$ - function showing a perfect agreement with the previous results in the particular case in which
the Renyi entropy becomes the von Neumann entropy.

\section{Scaling of entanglement}

The logarithmic scaling law that 
entanglement entropy obeys in the critical regime is a sign
of the conformal symmetry of the system. 
For second order phase transitions, the correlation length diverges and the 
system becomes scale invariant. This scaling symmetry gets enlarged 
to the conformal group \cite{Polyakov} which, in the case
of on-dimensional systems, allows for a very precise characterization
of the operator structure of the underlying theory.
The development of conformal field theory is a remarkable
achievement that we cannot present in this short review \cite{Ginsparg, Cardy}.

\subsection{One-dimensional systems}
For a conformal theory in 1+1 dimensions, 
the scaling behaviour of the entropy was 
proven to be logarithmic in Ref.\ \cite{HoLaWi}.
The general result reads
\be S_L\sim
\frac{c+\bar{c}}{6} \log_2 L \, ,
\label{Eq:scaling-entropy-CFT}
\ee 
where $c$ and $\bar c$ are the so
called central charges for the holomorphic and anti-holomorphic sectors
of the conformal field theory.
These central charges classify conformal field theories 
and are {\it universal} quantities which depend only on basic
properties of the system, like effective degrees of freedom of the theory,
symmetries or spatial dimensions,
and they are independent of the microscopic details
of the model.
For free bosons $c=1$, whereas  for free fermions $c=1/2$.

This result matches perfectly our geometric entropy computation
of the critical XX model. 
In this case, the central charge  $c=\bar c =1$ and the model
is seen to belong to the free boson universality class.

The previous result of Eq.\ (\ref{Eq:scaling-entropy-CFT}) was further elaborated
and extended
to finite systems, 
finite temperature and disjoint regions in Ref.
\cite{HoLaWi,CC04,CC09}.
For instance, the scaling of the entropy for a system with periodic boundary conditions 
reads
\begin{equation}
S_A\sim \frac{c}{3}\log\left(\frac{L}{\pi a}\sin\frac{\pi\ell}{L}\right)+c_1'\, ,
\end{equation}
whereas for the open boundary conditions case is
\begin{equation}
S_A=\frac{c}{6} \log \left(\frac{2L}{\pi a}\sin \frac{\pi\ell}{L}\right)+{\tilde c}'_1\,.
\end{equation}
In Ref.\ \cite{ZBFS06}, the scaling of the entropy of a conformal semi-infinite chain 
is presented.
In Ref.\ \cite{CL08}, conformal symmetry is further exploited 
and an analytical computation of the
distribution of eigenvalues of the reduced density 
matrix of a block in a one-dimensional systems is presented.

Let us finally mention that the scaling of entanglement
have been also studied for other entanglement measures by means of
conformal field theory. In particular, in Ref.\ \cite{EC05,OLEC06}, 
it is shown that the single copy entanglement scales as
\be
E_1(\rho_L)=\frac{c}{6} \log L- \frac{c}{6}  \frac{\pi^2}{\log L} + O(1/L) \, .
\ee
Note that entropy sub-leading corrections to scaling are suppressed
as $1/L$ whereas single copy entanglement suffers from $1/\log L$ modifications.
This makes the numerical approach to the latter more difficult.

\subsection{Conformal field theory and central charge}

We stated above that the central charge is a quantity
that characterizes the universality class of a quantum
phase transition. We also mentioned in the introduction
that a possible figure of merit for entanglement
could be constructed from correlation functions made
out some operator which is always present in any theory.
Let us now see that both ideas merge naturally.

In 1+1 dimensions, conformal field theories are
classified by the representations of the conformal group \cite{Ginsparg}. 
The operators of the theory fall into a structure of
highest weight operators and its descendants. Each highest weight
operator carries some specific scaling dimensions which dictates those
of its descendants. The operators close an algebra implemented into
the operator product expansion.  One operator is particularly
important: the energy-momentum tensor $T_{\mu\nu}$, which is a descendant of the identity. 
It is convenient
to introduce holomorphic and anti-holomorphic indices defined by the
combinations $T=T_{zz}$ and $\bar T=T_{\bar z \bar z}$ where $z=x^0+i
x^1$ and $\bar z=x^0-i x^1$. Denoting by $\ket{0}$ the vacuum state, the central charge of a conformal field
theory is associated to the coefficient of the correlator 
\be \langle
0\vert T(z) T(0)\vert 0\rangle = \frac{c}{2 z^4} \, ,
\ee 
and the analogous
result for $\bar c$ in terms of the correlator $\langle 0\vert \bar
T(z) \bar T(0)\vert 0\rangle$.  A conformal field theory is
characterized by its central charge, the scaling dimensions and
the coefficients of the operator product expansion. Furthermore,
unitary theories with $c<1$ only exist for discrete values of $c$ and
are called minimal models. The lowest lying theory corresponds to
$c=\frac{1}{2}$ and represents the universality class of a free
fermion. 

The central charge plays many roles in a conformal field theory.  It
was introduced above as the coefficient of a correlator of
energy-momentum tensors, which means that it is an observable. The
central charge also characterizes the response of a theory to a
modification of the background metric
where it is defined.  Specifically, the scale anomaly
associated to the lack of scale invariance produced by a non-trivial
background metric is 
\be 
\langle 0\vert T^\mu_\mu \ket{0} =
-\frac{c}{12} R \, ,\ee 
where $R$ is the curvature of the background
metric. This anomaly can also be seen as the emergence of a non-local
effective action when the field theory modes are integrated out in a
curved background.

Therefore, the central charge which appears as the coefficient
of the entanglement entropy is naturally related to the 
stress tensor, which is the operator that is guaranteed
to exist in any theory. It is also possible to derive
a direct relation between entropy and the trace of the
stress tensor as shown in the original Ref.\ \cite{HoLaWi}.

\subsection{Area law}

The conformal group does not constrain the structure
of the Hilbert space in spatial dimensions higher than
one as much as it does in one dimension. Actually,
the conformal group no longer brings an infinite number
of conserved charges (as it does in one dimension) but
becomes a finite group.

Nevertheless, 
a geometric argument establishes the scaling behaviour of
entropy. The basic idea goes as follows.
Let us consider a volume of spins (or any local degrees
of freedom) contained in a larger space. For theories
with local interactions, it is expected that entanglement
will be created between the degrees
of freedom that lie outside and inside the surface that encloses
the volume we are considering. It follows that the entropy
should naturally scale as an area law even if the model
displays a finite correlation length.

These arguments were put forward in the study of entanglement
entropy in quantum field theory as a possible source for black-hole entropy 
\cite{S86,Bombelli86,FPST94}.
Furthermore, the relation between the entropy and
the effective action in a curved background was developed
in Ref.\ \cite{KS94}. Let us mention these results. 
For general quantum field theories the entropy is
a divergent extensive quantity in more than one spatial
dimension obeying an area law
\be
S_L \sim c_1 \left(\frac{L}{\epsilon}\right) ^{d-1} \qquad\qquad  d>1 \, ,
\label{general-entropy}
\ee
where $L^d$ is the size of the volume, $\epsilon$ stands for an ultraviolet regulator
and the coefficient $c_1$ counts components of the field which is considered.
This coefficient is again found in the effective action on a gravitational
field and, thus, in the trace anomaly as a divergent term. A form for the former
can be found as
\be
\Gamma_{eff}=\int_{s_0}^\infty ds \frac{e^{-sm^2}}{s^{d/2}}
\left( \frac{c_0}{s}+c_1 R + c_{2F} s F+ c_{2G} s G+\ldots\right) ,
\label{general-entropy2}
\ee
where $s_0$ acts as a ultraviolet regulator, $R$ is the curvature,
$F$ the Weyl tensor and $G$ the Euler density. The main conceptual
result to be retained is that entropy measures a very basic
counting of degrees of freedom. Note that previous efforts to
make a general c-theorem are all base on $c_{2F}$ an $c_{2G}$,
not on $c_1$. In one spatial dimension, the effective action
has a unique structure proportional to the central charge.
That is, the central charge takes over all manifestations of
the trace anomaly, at variance with the separate roles
that appear in higher dimensions.

It is worth mentioning that computations done in massive
theories in any number of dimensions show that
$S_L(m\not= 0)-S_L(m=0)$ comes out to be ultraviolet finite \cite{Ka95}.
Actually, the ultraviolet cut-off cancels in the computation.
This is precisely what is needed to make the RG flow
meaningful in such a case. This comment hints at the non-trivial
issue about observability of the entropy. The standard prejudice
is that the leading area law coefficient is not observable since
it comes divided by a necessary ultraviolet cut-off. Yet, if such
a coefficient is also responsible for finite corrections, the situation
may not be as trivial. 

A review on methods to calculate the entanglement entropy for free fields
and some particular examples in two, three and more dimensions are presented
in Ref.\ \cite{CH09} in this special issue.
Further explicit computations of area law scaling of entropy in spin
and harmonic systems in higher dimensions can be found in 
Refs.\ \cite{RL06, CEPD06, BCS06, ECP08}. 
A quite remarkable result found in Ref.\ \cite{Wolf} is
the fact that certain fermionic systems may develop
logarithmic violations of the area law, while keeping local interactions.
This is somehow analogous of the logarithmic scaling
in one-dimension. The system is more correlated than
what is expected from pure geometrical arguments.
In this respect, the leading term in the scaling of the entropy for fermionic
systems was computed analytically assuming the Widom conjecture in Ref.\ \cite{GK06}.
This result was checked numerically for two critical fermionic models
in Ref.\ \cite{BCS06} finding a good agreement.

For other  steps into a description of systems
with two spatial dimensions in the framework of conformal field
theory see Refs.\ \cite{Fradkin06,RT06}. 
For a class of critical models in two spatial dimensions 
(including the quantum dimer model), it is found that 
        $ S(\rho_I) = 2 f_s (L/a) + c g \log(L/a)+O(1)$,
where $L$ is the length of the boundary area, 
$f_s$ is an area law coefficient that is interpreted as a {\it boundary
free energy}, and $g$ is a coefficient that depends on the 
geometric properties of the partition.
That is, in addition to a non-universal area law, 
one finds a universal logarithmically divergent correction. 
For a further discussion of steps towards a full theory of entanglement
entropies in $d+1$-dimensional conformal field theories,
see Refs.\ \cite{Fradkin06,RT06}. 

A particularly interesting issue is the holographic entanglement entropy 
that emerges from the AdS (anti-de-Sitter)/CFT correspondence.
The AdS/CFT correspondence is the conjectured equivalence 
between a quantum gravity theory defined on one space, 
and a quantum field theory without gravity defined on the conformal boundary of this space.
The entanglement entropy of a region of the boundary in the conformal field theory is then related
with the degrees of freedom of part of the AdS space in the dual gravity side.
In Refs.\ \cite{RT06, RT2006}, this relation is established explicitly and, in Ref.\ \cite{Takayanagi09}
in this special issue, the recent progress on this topic is presented.

Let us also mention the line of research that deals with
topological entropy. Some Hamiltonians produce states
such that a combination of geometric entropies exactly
cancels the dominant area law. Then, a topological 
entropy term characterizes the system \cite{KP06}.
This subject is nowadays a large field of research that
we cannot include in the present review.
In this respect, a review on 
the scaling of the entanglement entropy
of 2D quantum systems in a state with topological order
is presented in Ref.\ \cite{Fradkin09} in this special issue. 

\section{Other models}
We can find in the literature the computation of the scaling of the entanglement
entropy for other spin models.
In XY and XXZ models, this logarithmic scaling will confirm
the role of the underlying conformal symmetry. 
We shall also
discuss that in disordered systems, although the
conformal symmetry is lost for one particular realization of the disorder,
we recover the logarithmic scaling of the entropy with a different central charge of the 
corresponding homogeneous model,
if we do the average over all the possible realizations of the disorder.
We shall also study the scaling of entropy in systems where the
notion of geometric distance is lost.
This is the case of the Lipkin-Meshkov-Glick model, 
in which the logarithmic behaviour of the entropy will be due to the
equilibrium of a competition between the long range interactions, that try to increase
the entanglement, and the symmetries of the problem, that force the 
ground state to belong to a reduced subspace of the Hilbert space.
A different case are those systems composed of itinerant particles. In particular, we will
present the scaling of entropy of the Laughlin wave function.

\subsection{The XY model}
The XY model is defined as the XX model in Eq.\ (\ref{eq:XX-Hamiltonian}), 
adding an extra parameter
$\gamma$ that determines the degree of anisotropy of spin-spin interaction
in the XY plane. Its Hamiltonian reads
\be
H_{XY} = -\frac{1}{2}\!\sum_l \left( \frac{1\!+\!\gamma}{2} \sigma_l^x
\sigma_{l+1}^x +
 \frac{1\!-\!\gamma}{2} \sigma_l^y \sigma_{l+1}^y + \lambda\sigma_l^z  \right)\, ,
\label{eq:XY-model}
\ee
where, as in the previous section, $l$ labels the $N$ spins, $\sigma_l^{\mu}$ ($\mu=x,y,z$) 
are the Pauli matrices and $\lambda$ is the transverse magnetic field in the $z$ direction. 
This notation will be also followed for the rest of models that are going to be presented.

Notice that if $\gamma =0$ we recover the XX model, whereas
if $\gamma=1$, it becomes the quantum Ising model with a transverse magnetic field, 
with Hamiltonian
\be
H_{\rm Ising} = -\frac{1}{2}\!\sum_l
 \left( \sigma_l^x \sigma_{l+1}^x + \lambda\sigma_l^z  \right).
\label{eq:Ising-model}
\ee

The XY model was solved in detail in Ref.\ \cite{LRV04}. 
In order to do this, the previous works on spin chains
required to solve the XY Hamiltonian were reviewed. 
In concrete, the XY model without magnetic field 
was solved exactly in Ref.\ \cite{Ann}, 
the spectrum of the XY model with magnetic field was computed in Ref.\ \cite{Kat}, 
the correlation function for this model was obtained in Ref.\ \cite{Bar},
and the entropy $S_L$ was computed in Ref.\ \cite{vid02}.

Later, an extent analytical analysis of the entropy of XY spin 
chain was presented in Ref.\ \cite{Kor}. 
In this work, in a similar way as we have seen previously for the XX model, 
an analytical expression for the scaling of the entanglement 
entropy is determined for the XY model by means of Toeplitz determinants.

The XY model with $\gamma \ne 0$ is critical for $\lambda=1$. 
In this case, 
the entropy of a block scales as 
\be 
S_{XY}(L)=\frac{1}{6} \log_2 L + a(\gamma), 
\label{Eq:entropy-XYmodel}
\ee 
where $a(\gamma)$ is a function that only depends on $\gamma$.
This entanglement behaviour corresponds to
the scaling dictated by a conformal theory, Eq.\ (\ref{Eq:scaling-entropy-CFT}),
with central charge $c=1/2$. The XY model, therefore, falls into
the free fermion universality class.

In the non-critical case, that is for $\lambda\ne 1$, the entanglement entropy
saturates to a constant.

Let us mention that an exact relationship between 
the entropies of the XY model and the XX model
has been found recently \cite{IJ08}. 
Using this relation it is possible to translate known results between 
the two models and obtain, among others, the additive constant of 
the entropy of the critical homogeneous quantum Ising chain and 
the effective central charge of the random XY chain.

Finally, with respect to the particular case of the Ising model,
in Ref.\ \cite{CCD07}, the computation of the leading correction 
to the bipartite entanglement entropy at large sub-system size, 
in integrable quantum field theories with diagonal scattering matrices is presented. 
This result is used to compute the exact value of the saturation in the Ising model 
and showed it to be in good agreement with numerical results.
This work is reviewed in detail in Ref.\ \cite{CD09} in this special issue.

\subsection{The XXZ model}
The XXZ model consists of a chain of $N$ spins with nearest 
neighbour interactions and an external magnetic field.
Its Hamiltonian is given by,
\be
\label{eq:XXZ}
H_{XXZ} = \sum_l \left(\frac{1}{2}[\sigma^x_l\sigma^x_{l+1} + \sigma^y_l\sigma^y_{l+1} + 
\Delta \sigma^z_l\sigma^z_{l+1}]  +\lambda  \sigma^z_l \right),
\ee
where $\Delta$ is a parameter that controls the anisotropy in the $z$ direction. 

As it happened for the $\gamma$ parameter of the XY model, 
the $\Delta$ parameter of the XXZ model has two particular interesting values.
If $\Delta=0$, we trivially recover the XX model, and
if $\Delta=1$, the system becomes the XXX model that has a fully isotropic interaction
\be
\label{eq:XXX}
H_{XXX} = \sum_l \left(\frac{1}{2}[\sigma^x_l\sigma^x_{l+1} + \sigma^y_l\sigma^y_{l+1} + \sigma^z_l\sigma^z_{l+1}]  
+\lambda  \sigma^z_l \right).
\ee

The XXZ model can be solved analytically by means of the Bethe Ansatz technique \cite{Bethe31}.
Bethe Ansatz takes profit of the two symmetries of the system to find its eigenstates.
The first symmetry is the rotational symmetry respect to the $z$ axis.
It implies that the z-component of the  total spin, $S_z=1/2\sum_l \sigma_l^z$, 
must be conserved
and, therefore, the Hamiltonian must be diagonal in boxes of constant $S_z$.
The other symmetry is the translational invariance, that allow us to diagonalize
these boxes using a kind of generalized Fourier transform.
Once the ground state is obtained, the correlation functions
can be computed in terms of certain determinants (see Ref.\ \cite{BIK93}).
This model is qualitatively different from the XY, since it presents
a point of non-analyticity of the ground state energy for finite systems.
In the XY model, the level crossing between the ground state and the
first excited state only occurs in the thermodynamic limit.
In this case, instead, the terms of the Hamiltonian
$\sigma^x_l\sigma^x_{l+1} + \sigma^y_l\sigma^y_{l+1}$, 
$\sigma^z_l\sigma^z_{l+1}$ and $\sigma^z_l$ commute and
are independent of $\Delta$ and $\lambda$, 
which implies that there will be an actual level crossing.

Both the isotropic case and the anisotropic one
for $\lambda=0$ are solved in Refs.\ \cite{Takahashi99, LRV04}.
The phases of the system are found to be:
\begin{itemize}
\item In the XXX model, Eq.\ (\ref{eq:XXX}), there are two limit behaviours.
On one hand, when $|\lambda| > 2$ the system is gapped and 
it is in a product state in which
all spins point at the direction of the magnetic field (ferromagnetic phase).
On the other hand, for $\lambda=0$ the magnetization is zero and 
the system is in a entangled state (anti-ferromagnetic case). 
In the interval between these
two cases $2>\lambda\ge 0$ the system is gap-less and, therefore, critical.

\item With respect to the anisotropic case with magnetic field equal to zero, 
the system shows a gap-less phase in the
$1 \ge \Delta \ge -1$ interval. Outside this interval, there is a gap between 
the ground and the first excited states. 
These two phases are separated by two phase transitions in $\Delta=1$ and
$\Delta=-1$. 
The first one is a Kosterlitz-Thouless phase transition, 
while the second one is of first order.
\end{itemize}

The scaling of the entanglement entropy is presented in Refs.\ \cite{LRV04,EER09}.
These numerical results show that the entanglement entropy
behaviour converges to a logarithmic scaling as the size of
the system increases, if the system is critical.
On the contrary, if the system is not in a critical phase,
the entropy saturates to a constant value.
In particular, in the isotropic model without magnetic field, 
the entropy scales as
\be
 S_L\sim \frac{1}{3} \log_2 L\ ,
\ee
which means that the XXX model with $\lambda=0$ has central 
charge $c=1$ and falls into the universality class of a free
boson.

Finally, let us mention that, recently, analytic expressions for 
reduced density matrices, several correlation functions 
and the entanglement entropy of small blocks (up to 6 spins) 
have been found for the XXZ model with $\Delta=1/2$ 
(see Refs.\ \cite{SS07} and \cite{NCC09}).

\subsection{Disordered models}
So far, we have only considered translational invariant systems. 
This symmetry plus the scaling invariance at
a critical point produces the conformal symmetry that implies
universal properties of the scaling of entanglement.
Nevertheless, natural systems exhibit a certain amount of {\it disorder}
due to impurities and imperfections of the system.
This disorder breaks the translational symmetry 
and we wonder what happens with the scaling of the entropy
taking into account that the conformal invariance is lost.

This question was addressed 
in Ref.\ \cite{RefaelMoore04}.
They computed analytically the block entropy for the 
Heisenberg, XX and quantum Ising models with random nearest-neighbour couplings
under the hypothesis of strong disorder by means of the real space renormalization
group technique.
This approach 
was introduced 
in Ref.\ \cite{Fisher94} and was generalized
in Ref.\ \cite{DM80}.

This strong disorder
hypothesis assumes that if one takes the strongest coupling of
the chain, its neighbours are much weaker than it. Thus,
it is possible to diagonalize this strongest bond
independently of the rest of the system, project the
system onto the ground state of this subspace (a singlet 
for the previous models) and 
then perform perturbation theory respect to the neighbour couplings.
The final result is that two sites have been eliminated
and the Hamiltonian energy scale has been reduced.
This process can be iterated
until we arrive at the ground state of the system
which is a random
singlet phase, that is to say, a set of
singlets connected randomly and for arbitrarily long distances.

Notice that although this method is not correct when applied
to a system with weak disorder, it becomes
asymptotically correct at large distances \cite{Fisher94}.

For a particular realization of the disorder, the translational symmetry
of the system is broken and, therefore, the conformal symmetry too.
Hence, the scaling of the entanglement entropy of this realization of the disorder
will not be logarithmic, but fluctuating.

In Ref.\ \cite{RefaelMoore04}, it was shown
that although the conformal symmetry 
is broken, 
if we take the average over all the realizations of the disorder
the entropy keeps scaling logarithmically
with an effective central charged $\tilde{c}=c \ln 2$, 
where $c$ is the central charge for the same model but without disorder.
This result has been further checked numerically both
for the XX model in Ref.\ \cite{Laflorencie05}
and for the Heisenberg model in Ref.\ \cite{DMCF06}.

In Ref.\ \cite{Laflorencie05}, 
the disordered XX spin-$\frac{1}{2}$ chain
with periodic boundary conditions and 
positive random spin couplings chosen in a flat uniform
distribution within the interval $[0,1]$ was studied. 
The magnetic field was set to zero.
It was shown that for a block large enough (larger than 20 spins),
the entropy scales logarithmically according to \cite{RefaelMoore04},
using around $10^4$ samples for 
$N=500,~1000,~2000$ and $2\times 10^4$ samples for $100\le N \le 400$,
in order to do the average over all the possible realizations of the disorder.

The same result was shown for the Heisenberg model in Ref.\ \cite{DMCF06}.
In this work,  a uniform distribution in the interval $[0,1]$ for the couplings between the spins
was also chosen.
For a system of $N=50$ and after averaging the entanglement entropy over $10^4$ different
configurations of disorder, the logarithmic scaling of the entropy
with an effective central charge $\tilde{c}=c \ln 2$ is recovered.
Let us point out that these one dimensional systems are particular cases of 
chains of quantum group (or q-deformed) spins studied in Ref.\ \cite{BK07}.
It is also interesting to mention that this robustness of the entanglement scaling
respect to the disorder is not kept for other models like the Bose-Hubbard model (see Ref.\ \cite{FC08}).

In the case of higher dimensions,	
the scaling of the entanglement entropy in the
2D random Ising model was studied in Refs.\ \cite{LIR07,YSH08}.
In particular, in Ref.\ \cite{YSH08}, 
the entanglement entropy of a $L\times L$ region located
in the centre of a square lattice which is governed by the Hamiltonian
\begin{equation}
H = -\sum_{\langle i,j\rangle} J_{ij} \sigma^z_i \sigma^z_j - \sum_i \lambda_i
\sigma^x_i \, ,
\label{Eq:Ising2D}
\end{equation}
was computed.
The Ising couplings $J_{ij}$ and the transverse magnetic fields $\lambda_i$ take
random values given by the uniform probability distributions in the intervals
$[0,1]$ and $[ 0,\lambda_0 ]$ respectively.
By means of a generalized version for 2 dimensions of the real space renormalization group,
it was found that the critical field is at $\lambda^c_0 = 5.37\pm0.03$, 
and for both critical and
non-critical $\lambda_0$ the entropy scaling fulfils the area law: $S(L)\sim L$ in
the leading term.

Let us mention some disordered spin systems have also been studied
from the fidelity susceptibility point of view in Ref.\ \cite{Zanardi-disorder}.
Finally, it is interesting to point out that, in other systems, 
the translational invariance is not broken 
by means of random couplings but due to a quantum impurity or a physical boundary.
The behaviour of the entanglement entropy in this kind of systems is reviewed 
in Ref.\ \cite{ALS09} in this special issue.

\subsection{The Lipkin-Meshkov-Glick model}
The Lipkin-Meshkov-Glick model was proposed in Ref.\ \cite{LMG65}.
Unlike the previous models we have considered, where the spins had short range interactions,
in the LMG model, each spin interacts with all the spins of the system
with the same coupling strength. 
This system is described by the Hamiltonian
\be
H_{LMG}=- \frac{1}{N}\sum_{i<j} \left( \sigma_i^x \sigma_j^x + \gamma
\sigma_i^y \sigma_j^y
\right) - \lambda \sum_i \sigma_i^z \, .
\label{Eq:LMG}
\ee
Notice the apparent similarity between this model and the XY model in Eq.\ (\ref{eq:XY-model}).
The essential difference between them is that 
while in the XY Hamiltonian the interaction only takes place between nearest neighbours,
in the LMG model, all spins interact among themselves.
This highly symmetric interaction pattern 
forces the loss of the notion of geometry, since there is no
distance between the spins. This implies that it no longer makes sense to define
a block of $L$ spins as a set of $L$ contiguous spins or to study decays of the 
correlations between two spins. 

As in previous cases, our aim is to study the scaling
properties of the entanglement entropy for the ground
state reduced density matrix of a block of $L$ spins
respect to the rest of $N_L$ spins. We face a somewhat
contradictory situation.
On one hand, we expect that the non-local connectivity of
the interactions would produced a ground state more entangled than
those that emerge from nearest neighbour interaction models.
On the other hand, the symmetry of the Hamiltonian implies
that all the spins must be indistinguishable in the ground state,
therefore, it must belong to a symmetric subspace,
which restricts its entanglement. The explicit computation
will clarify this issue.

The Hamiltonian (\ref{Eq:LMG}) can
be written in terms of the total spin operators $S_{\alpha}= \sum_i
\sigma_{\alpha}^i / 2$ as
\begin{eqnarray}
H&=& - \frac{1}{N} (1+\gamma) \left({\bf S}^2-S_z^2 -N/2 \right)
-2\lambda S_z  \nonumber \\
&&- \frac{1}{2 N} (1-\gamma)\left(S_+^2+S_-^2\right) \ , \label{Eq:LMG2}
\end{eqnarray}
where $S_\pm$ are the ladder angular momentum operators.
In Eq.\ (\ref{Eq:LMG2}), we realize that $[{\bf S}^2,H]=0$ and, therefore, 
we can diagonalize the Hamiltonian in boxes of constant $S$.
From Eq.\ (\ref{Eq:LMG2}), it is easy to see that the ground state
must belong to the subspace of $S=N/2$.
Then, we have to span this subspace in terms of a basis $\ket{N/2,M}$
fully symmetric under the permutation group and eigenstates
 of  ${\bf S}^2$ and $S_z$. 
These states $\ket{N/2,M}$ are called Dicke states.

Notice that the restricted subspace 
where the ground state must live due to the symmetries
of the Hamiltonian
will limit the scaling of the entanglement entropy 
of a block of $L$ spins with respect to the remaining $N_L$.
As the ground
state reduced density matrix is spanned by the set of $(L+1)$
Dicke states, the entropy of entanglement obeys the constrain $S_{L,N} \le
{\rm log}_2 (L+1)$ for all $L$ and $N$, where the upper bound corresponds to
the entropy of the maximally mixed state $\rho_{L,N} = \nbOne/(L+1)$ in
the Dicke basis. This
argument implies that entanglement, as measured by the Von Neumann
entropy, cannot grow faster than the typical logarithmic scaling
observed in the previous cases.

Both the ground state and the entanglement entropy were computed for the LMG model
in Ref.\ \cite{LORV05}.
For the isotropic case ($\gamma=1$) and in the thermodynamic limit ($N,L \gg 1$),
 $H_{LMG}$
is diagonal in the Dicke basis. Then, for $\lambda \ge 1$, the
entanglement entropy is strictly zero since the ground state is in a fully polarized 
product state. Instead, if $1 > \lambda \ge 0$, we recover the logarithmic scaling
of the entropy,
\begin{equation}
S_{L,N} (\lambda,\gamma=1) \sim \frac{1}{2} \log{[L(N-L)/N ]}.
\label{Eq:entropy-LMGisotropic}
\end{equation}
Although the kind of scaling does not depend on the strength of the magnetic field,
its absolute value is smaller for weaker magnetic fields, according to the equation
\begin{equation}
S_{L,N} (\lambda,\gamma=1) -S_{L,N} (\lambda=0,\gamma=1) \sim \frac{1}{2} \log {\left(1-\lambda^2 \right)},
\label{eq:Shiso}
\end{equation}
and thus diverges, at fixed $L$ and $N$, in the limit $h\rightarrow 1^-$.

In the anisotropic case, we can study the limits
of very strong and very weak magnetic fields. 
On one hand, when $\lambda \to \infty$, the GS is in the product state
$\prod_i \ket{\uparrow}_i$ and therefore is not entangled. 
In the thermodynamic limit, this state is also the ground state just for  $\lambda >1$.
On the other hand, for $\lambda \to 0$ the entanglement entropy saturates and goes to a constant that
depends on $\gamma$. In the particular case of $\gamma=0$, the ground state is degenerate
and lives in the subspace generated by the states $\prod_i \ket{\to}_i$ and $\prod_i \ket{\gets}_i$,
where $\ket{\to}$ and $\ket{\gets}$ are the eigenstates of the $\sigma^x$ operator.
In practice, this degeneration would be broken by any perturbation of the environment.

These two different phases suggests the existence of a quantum phase transition between $\lambda \gg 1$ and
$\lambda \ll 1$. In particular, it has, numerically, been checked in Ref.\ \cite{LORV05} that, in the 
thermodynamic limit, the entanglement entropy displays a logarithmic divergence around $\lambda_c=1$ 
according to the law 
\begin{equation}
S_{L,N} (\lambda,\gamma) \sim - \log{|1-\lambda|}.
\label{eq:Shani}
\end{equation}
Indeed, it is shown that at $\lambda=1$ the entropy scales logarithmically
with a coefficient that depends on $\gamma$. However, in the thermodynamic limit,
this coefficient is independent of $\gamma$ and takes a value closed to 1/3.
In Ref.\ \cite{BDV06}, the previous relation, Eq.\ \ref{eq:Shani}, is computed analytically obtaining
the same result and fixing the coefficient to $1/3$. 
In this same work, the finite size corrections to
the scaling of entropy are also studied.

Although the behaviour of entanglement is very similar to the XY model,
that is to say, it scales logarithmically in the critical point and saturates to a constant 
in the non-critical phase, the reasons of these scaling laws are different.
In the XY model, entanglement is limited by the facts that interactions are local
and the system is translationally invariant. At the critical point, the correlation
length becomes infinite, the system
is conformal symmetric, and the logarithmic scaling of the entanglement entropy
appears as a manifestation of this symmetry.
Instead, in the LMG model, the long range interactions should allow 
for larger correlations, that is, larger entanglement. Nevertheless, the symmetries of the system 
restrict the subspace where the GS must belong and, therefore, the scaling law of entanglement.
The final result is the same logarithmic scaling law but which has nothing to do with
any underlying conformal symmetry.

Finally, let us mention that other analytical calculations of
the spectrum of the LMG model both in the thermodynamic limit and finite size case
have appeared recently \cite{RVM08}. 
Moreover,
the entanglement entropy for general free bosonic two-mode models is presented in Ref.\ 
\cite{VDB07}. In particular, a complete classification of the possible 
scaling behaviours for the entanglement entropy in the related collective models as the LMG, 
the Dicke model, or the Lieb-Mattis model is obtained.

\subsection{Particle entanglement}

In a similar way than LMG model, where the notion of distance was lost, one
can try to compute the entanglement entropy in systems of moving fermions and bosons.
In such itinerant systems, as the particles are indistinguishable, 
moving and partially de-localized, 
it is not obvious to define the geometric entropy. 

What we, indeed, can compute is the von Neumann entropy for any 
subset of particles for a system of $N$ indistinguishable 
particles in the state $\Psi(r_1,\ldots,r_n)$. 
Notice that, in this case, 
this von Neumann entropy cannot be interpreted as 
the number of distillable EPR pairs. 
Due to the symmetrization, it is impossible to associate a 
label with the particles and perform the appropriate distillation operations.
This is a subtle difference respect to the LMG model.

A particular interesting physical system is 
the Fractional Quantum Hall Effect (FQHE) \cite{Tsui82}. 
Although 
a complete understanding of it is still missing, 
it is commonly believed that the interactions between 
the particles are essentially responsible for the 
strange states of matter that the 2D electron gas shows
at some particular values of the transverse magnetic field. 
These states would present a new kind of order called topological order 
and their quasi-particle excitations are neither bosons nor fermions, 
but anyons, that is to say, quasi-particles with any-statistics \cite{Wen}.
In this respect, in 1983,
Laughlin proposed an Ansatz for the wave function of the ground state of the system \cite{L83}. 
This wave function is defined by
\begin{equation}
\label{eq:Laughlin-def}
\Psi_L^{(m)}(z_1,\ldots,z_n)\sim
\prod_{i<j}(z_i-z_j)^{m} 
e^{-\sum_{i=1}^{N}\vert z_i \vert^2/2} \ ,
\end{equation} 
where $z_j=x_j+\textrm{i}y_j, j=1,\ldots,n$ 
stands for the position of the $j$-th particle.
It describes fractional quantum Hall 
state at a filling fraction $\nu=1/m$, where $m$ is an integer number.

In particular, in Ref.\ \cite{ILO07}, the entanglement entropy of $k$ particles
respect the rest of the system is computed for the Laughlin wave function
with filling fraction one
\begin{equation}
  S_{n,k} = \log_2\binom{n}{k}\, .
\label{eq:entanglement-entropy-laughlin}
\end{equation}
Notice that, in this case, although the state also belongs to a completely anti-symmetric subspace,
the entanglement entropy of half a system grows linearly with the number of particles. 

In Ref.\ \cite{ZHS08}, these ideas are extended, and the {\sl particle entanglement},
defined as the entanglement between two subsets of particles making up the system, is studied. 
The general structure of particle entanglement in many-fermion ground states, 
analogous to the {\sl area law} for the more usually studied entanglement between spatial regions, 
are also formulated, and the 
basic properties of particle entanglement are uncovered by considering relatively simple itinerant models.
All these ideas are widely reviewed in Ref.\ \cite{HZS09} in this special issue.

%
%
\section{Renormalization of Entanglement}

A natural question arising within the study of entanglement
in quantum system is how entanglement evolves along Renormalization
Group (RG) trajectories. We shall now address this issue
discussing first the RG of quantum states and, then, the study of
particular systems.

\subsection{Renormalization of quantum states}
It is customary to present RG transformations on Hamiltonians
or observables. In general, a Hamiltonian is described by a
set of coupling constants times operators $H=\sum_i g^i {\cal O}_i$.
This set of operators may be infinite, including relevant,
marginal and irrelevant operators or, as in the case of renormalizable
quantum field theories in the continuum, it may reduce to a finite set of relevant
and marginal operators. Then, upon coarse graining of short-distances
and an adequate rescaling, the system is described by a new
set of coupling constants. So to speak, the operator algebra
acts as a basis. The concept of RG trajectory corresponds to
analysing observables along the flow $\frac{\rm d}{{\rm d}t} =-\beta_i\frac{\rm d}{{\rm d}g^i}$,
where the beta functions correspond to $\beta_i\equiv \frac{\rm dg^i}{{\rm d}t}$ are related to the change of
the coupling constants as the coarse graining proceeds.

Yet, RG transformations can be understood as an action on any
quantum state, regardless of its relation to any Hamiltonian \cite{VCLRW05}.
Coarse graining is independent of any dynamics. This RG procedure
on quantum states is not presented as common lore since
explicit knowledge of {\sl e. g.} the ground state of a system 
is not available in general. Let us address this issue.

The basic idea to perform RG on states is to produce a
coarse graining of short-distance degrees of freedom,
followed by a clever choice of local basis to retain
the long-distance information which is retained in an optimal way.
Let us take a quantum state
$\psi_0$ and determine its RG transformed , $\psi_0'$, as follows. 
We pairwise group the sites in the
system and define a coarse-graining transformation for every pair
of local $d$-dimensional basis states, {\sl e.g.} for the sites $2j$ and $2j+1$,
as $\vert p\rangle_{2j}\vert q\rangle_{2j+1}= \vert pq\rangle_j $.
This transformation yields $\psi_0\to \psi$. Then we have
$\psi_0'=U\otimes \ldots \otimes U |\psi\rangle$, where the
$d^2\times d^2$ unitary matrix $U$ performs the change of
representative in the coarse--grained space. Note that the
matrix $U$ is non-local as seen from the $2j$ and $2j+1$ sites.
Some local information is now washed out, while preserving all the
quantum correlations relating the coarse--grained block to other
ones.

Operators also get coarse--grained along the above transformation.
Take for instance an operator acting on one local Hilbert space,
{\sl e.g.} $O_{2j}$. Expectation values must remain unchanged,
 \be
 \langle \psi_0\vert O_{2j}\vert \psi_0\rangle
 = \langle \psi_0'\vert O'_{j}\vert
 \psi_0'\rangle \, ,
 \ee
which leads to
 \be O'_l=U (O_{2j}\otimes I_{2j+1})U^{\dagger} \, ,
 \ee
where $I$ is the identity matrix. To complete a RG
transformation we simply need to rescale distances, i.e., to
double the lattice spacing.

This analysis can be made completely explicit  in the 
case of states which are described as a matrix product state
\cite{VCLRW05}.
There, the above transformation amount to a flow on
the matrices that represent the state. In turn, a flow
related to the transfer matrix can be computed. Explicit 
irreversibility of RG flows and the characterization of
critical points followed from the flow on this transfer matrix.

\subsection{Irreversibility of RG flows}

We may as well return to the standard construction
of RG transformations on Hamiltonians and perform
a detailed study in some particular case.
For instance, we may consider the quantum Ising model
in a transverse field $\lambda$. It is known that 
the parameter $\lambda$ provides a relevant deformation
of the model, departing form its critical value $\lambda^*=1$.
For instance, the departure that makes $\lambda>1$
get larger and larger corresponds to the increase of the
mass of the underlying fermionic description.

\begin{figure}[!ht]
\scalebox{0.6}{\includegraphics{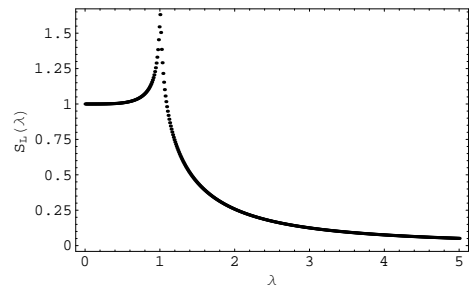}}
\caption{Entropy of entanglement is shown to decrease monotonically
along the RG trajectory that takes the external magnetic field 
$\lambda$ away from its critical 
value $\lambda^*=1$. Towards the left the flow takes the system
to a GHZ-line state whereas, towards the right, the system is
a product state.}
\label{fig:entropy-renormalization}
\end{figure}

An analysis of this RG trajectory can be illustrated
using Fig.\ \ref{fig:entropy-renormalization}
( see Refs.\ \cite{LLRV05,CC04}).
This result shows that RG trajectories are monotonically
irreversible as dictated by the c-theorem in 1+1 dimensions.
Furthermore, it can be seen that the ground state obeys
majorization relations. That is, irreversibility is
orchestrated at a very refined level, since the
reshuffling of the ground state obeys an exponential
set of ordering relations \cite{RL06}.

Irreversibility for the entanglement entropy should then
be obtained as a fundamental theorem, equivalent to the
c-theorem which is usually formulated in terms of the
stress tensor. This was indeed done in Ref.\ \cite{CH04}.
Once, the relation between entropy and the properties
of the stress tensor are made apparent.

%
%
\section{Dynamics of Entanglement}
So far, we have studied the properties of entanglement entropy of the ground state
of the system.
Next, we would like to analyse
how entanglement evolves in time when the system is prepared 
in a state that is {\em not} an eigenstate of 
the Hamiltonian.

\subsection{Time evolution of the block entanglement entropy}

In Ref.\ \cite{CC05}, 
the time evolution of the entropy of entanglement of a block of $L$ spins
in a one-dimensional system is studied. 
It is considered a system prepared in a pure state $|\psi_0\rangle$,
which corresponds to an eigenstate of $H(\lambda_0)$ with 
$\lambda_0\neq \lambda$. Then, for example,
at time $t=0$, the parameter is suddenly quenched from 
$\lambda_0$ to $\lambda$. In general, $|\psi_0\rangle$ 
will not be an eigenstate of $H(\lambda)$, and thus the system will evolve 
according to the equations of motion given by $H(\lambda)$.
In this work, two computations are performed: one based on conformal field theory
and the other on a particular solvable spin model, the Ising model.
In the first case, 
the path integral formulation and the CFT are used in order to 
calculate the time evolution of the entanglement entropy of a high energy
state of the system which is not an eigenstate. Then, one has to assume that
the Hamiltonian is critical in order to make the theory conformally invariant.
Instead, in the Ising model case, it is possible to perform calculations 
starting from a variety of initial
states, considering both critical and non-critical regimes.

In both calculations, 
the entanglement entropy increases linearly with
time $t$ (after transients die away in the lattice case), up to
$t^*=L/2$, in units where the maximum propagation speed of excitations
is taken to be unity. For $t\gg t^*$, $S_L(t) \sim L$ saturates at an
asymptotic value. This behaviour can be summarized in the following
equation:
\begin{equation}
 	S_{L}(t) \sim  
        \left\{ 
          \begin{array}{cc}
            t    &   \;\;\;\;\;  t \le t^* \\
            L    &   \;\;\;\;\;  t \ge  t^*
          \end{array}
          \right. \, .
\label{Eq:entanglement-dynamics}
\end{equation}

This behaviour of the entanglement entropy
has been checked in several lattice models both analytically and numerically
\cite{DMCF06, FCalabrese08, CDRZ07, SWVC08, LK08}.
In particular, in Ref.\ \cite{FCalabrese08}, the previous results are provided 
analytically using Toeplitz matrix representation and 
multidimensional phase methods for the XY model and considering large blocks.

\begin{figure}[ht]
\begin{center}
\includegraphics[width=0.50\linewidth]{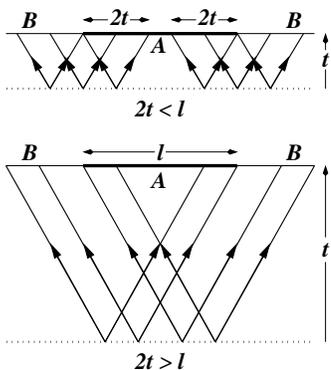}
\end{center}
\caption{Schematic representation of the dynamics of block entropy. Entangled 
particles are emitted from the region $A$, they will contribute to the block 
entropy as long as one of the two particles ends in the region $B$ 
[from \protect\cite{CC05}].}
\label{fig:entprod2}
\end{figure}

In Ref.\ \cite{CC05}, a simple interpretation of this behaviour is proposed in terms of 
quasi-particles excitations emitted from the initial state at $t=0$ and freely 
propagating with velocity $v \le 1$.
The idea is that at, $t=0$ and at many points of the chain, a pair of entangled quasi-particles begin
to propagate in opposite directions at some constant velocity $v$ that we will consider $1$ for simplicity
(see Fig.\ref{fig:entprod2}). The entanglement between the block of $L$ spins and
the rest of the system at an arbitrary time is given by the number of pairs that have one 
quasi-particle in the block while the other is outside. 
Thus, the entanglement entropy increases linearly with time
until it saturates when 
the excitations that started in the middle of the block arrive at its boundary.

All the previous results are explained in detail in the Ref.\ \cite{CC09},
where, apart from quantum quenches, a general conformal field theory 
approach to entanglement entropy is reviewed.

%
%

Let us point out that this increase of the entanglement
entropy is unrelated to the second law of thermodynamics. 
Entanglement entropy can decrease or even 
oscillate in standard time evolution.

Let us also mention that in Ref.\ \cite{DMCF06}
the dynamics of entanglement was analysed
for disordered systems, 
{\sl i. e.} when the couplings between the spins take random values.
In particular, the XXZ model with the couplings between the spins
following a uniform distribution in the interval $[0, 1]$ was studied.
It turns out that, in the presence of disorder, the entanglement entropy does not increase
linearly but logarithmically. 
This logarithmic behaviour 
does not follow from an extension of the argument for the clean case assuming
a diffusive propagation of the excitations,
but it requires some kind of entanglement localization.
This behaviour is also observed in Ref.\ \cite{BO07} where the 
propagation of information through the disordered XY model is studied. 
In particular, both classical and quantum correlations 
are exponentially suppressed outside of an effective light-cone 
whose radius grows at most logarithmically with time.

\subsection{Bounds for time evolution of the block entropy}
All these results are compatible with the rigorous bounds
found in Refs.\ \cite{BHV06, EO06} by means of the Lieb-Robinson bound \cite{LR72} and
its generalizations presented  
in Refs.\ \cite{Hastings}.

The Lieb-Robinson bound states that the operator norm
of the commutator of two operators $O_A$ and $O_B$
that act on different regions $A,B$ 
of a spin network with local interactions, $h_{ij}(t)$,
and in different times verifies
\be
\left\|\, \left[ O_A(t), O_B(0)\right]\, \right\|
\leq c N_{min}  \|O_A\|\, \|O_B\| \,
e^{-\frac{L-v|t|}{\xi}},
\label{Eq:Lieb-Robinson}
\ee
where $L$ is the distance between $A$ and $B$ 
(the number of edges in the shortest path connecting $A$ and $B$), 
$N_{min}=\min\{ |A|, |B|\}$ is the number of spins in the smallest of $A$ and $B$,
while $c,v,\xi>0$ are  constants depending only on
$g=\max_{(i,j)\in E} \max_t \|h_{ij}(t)\|$ and the architecture of the
spin lattice.

Thus, the Lieb Robinson bound, Eq.\ (\ref{Eq:Lieb-Robinson}), tells us that
the norm of the commutator of two operators at different times
is exponentially small outside the light-cone given by the
velocity $v$ that we can understand like the speed of sound.
Notice that, by dimensional analysis, 
this velocity must be proportional to the energy scale $g$.
It is interesting to point out that 
this result is also valid for the case of fermions or 
local Hamiltonians with exponentially
decaying interactions.

In Ref.\ \cite{BHV06}, it is shown, using the Lieb-Robinson bound and its generalizations \cite{Hastings}, 
that correlations
and information are propagated at a finite velocity in a spin network with
nearest-neighbour interactions. 
This is a non-trivial result since in non-relativistic quantum
mechanics there doesn't exist the notion of a light-cone, i. e. 
local operations could be used, in principle, to send information at
long distances in arbitrary short times.

Moreover, it is quantified the entanglement entropy that can be generated
per unit of time between a block of spins and the rest of the system.
In particular, it is found that
\be
S_L(t)-S_L(0)\le c^* g P t
\ee
where $c^*\simeq 1.9$ is a constant 
and $P$ is the perimeter of the block.
Finally, let us mention that all these results are complemented in Ref.\ \cite{EO06}.

\subsection{Long range interactions}
The Lieb-Robinson bound is only valid for short range interactions. 
Then, it is interesting to study how does entanglement evolve in
systems with long range interactions. 
This question is addressed in Ref.\ \cite{DHHLB05}.

In general, systems with long range interactions are numerically 
intractable since, in them, the entanglement entropy scales with the 
volume $S_L \sim L$.
Nevertheless, in Ref.\ \cite{DHHLB05},  the interactions are restricted
to Ising-type which allows to study both the static and the 
dynamical entanglement properties of the system.

It is considered a lattice composed by N spins that interact according to
the Hamiltonian,
\be
H=\sum_{k < l} f(k,l) \frac{1}{4}(\nbOne-\sigma_z^{(k)})\otimes(\nbOne-\sigma_z^{(l)}) \, ,
\label{Eq:InteractionHamiltonian}
\ee
where the coefficients $f(k,l)$, that describe the strength of the interaction between
the spins $l$ and $k$, obey a distance law, that is to say, $f(k,l)=f(\parallel k-l \parallel)$. 

It is assumed that the initial state is a product state of all spins
pointing to the x-direction $|\Psi_0\rangle = |\to\rangle^{\otimes N}$.
In order to perform the time evolution of this state, a description
in terms of Valence Bond Solids (VBS) is used (see Ref.\ \cite{VC04}).
With this method, it is possible to calculate the reduced density operator of few particles
for large systems (the computational time grows linearly with the whole size of the
system but exponentially with the size of the block).

In concrete, it is studied for some fixed time $t$ the scaling properties of entanglement
of a system with algebraically decaying interactions $f(k,l)=\parallel k-l\parallel^{-\alpha}$.
It turns out that for $\alpha\le 1/2$ (strong long-range interactions) 
the entanglement grows unbounded and the correlations do not practically decay,
while for $\alpha > 1$ the system contains a bounded amount of entanglement and
the correlations decay algebraically. 

The dynamics of entanglement are also studied. 
In the limit of an infinite chain, the entanglement entropy of any block
saturates for large times ($t \to \infty$) to its maximal value $S_L=L$ in a similar
way as in Eq.\ (\ref{Eq:entanglement-dynamics}).

\section{Entanglement along quantum computation}

It is known that slightly entangled quantum systems can be simulated efficiently
in a classical computer \cite{Vi03,Latorre07}.
This implies that any quantum algorithm that would exponentially accelerate a classical computation 
must create, at some point, a highly entangled state. Otherwise, the quantum algorithm
could be simulated efficiently in a classical computer.

Next, we want to briefly study how the entanglement evolves along a computation.
In order to do this, we will consider the three most common paradigms of quantum computation:
quantum circuits, adiabatic quantum computation, and one way quantum computing.

\subsection{Quantum circuits}
A quantum circuit is a sequence of unitary transformations (quantum gates)
on a register of qubits (see Ref.\ \cite{NC00} for a pedagogical introduction).
An efficient quantum circuit is characterized by the fact that the
number of elementary gates that form it only
scales polynomially in the number of qubits of the register.

The study of entanglement along a quantum circuit was
addressed in Refs.\ \cite{LM02} and \cite{OLM02} by means of majorization theory.
In these works the introduction of entanglement in Shor's 
algorithm and the Grover's algorithm were analysed respectively.

Let us remind the concept of Majorization relations, which is
a more refined measure of ordering of probability distributions than
the usual entropy one.
We say that a probability distribution $\{p_i\}$ majorizes 
another probability distribution $\{q_i\}$ (written as
$\vec{p}\prec \vec{q}$) if, and only if, 
\begin{equation}
\sum_{i=1}^k p_i \leq \sum_{i=1}^k q_i \qquad k = 1 \ldots d-1 \ ,
\label{maj2}
\end{equation}
where $d=2^N$ is the number of possible outcomes and it will correspond to the
dimension of the Hilbert space.

These Majorization relations can be related to quantum circuits in the following
way: let $|\psi_m \rangle $ be the pure state representing the
register in a quantum computer in the computational basis 
at an operating stage labeled by
$m = 0, 1\ldots M-1$, where $M$ is the total number of steps of
the algorithm. We can naturally associate a set of sorted
probabilities $p^{(m)}_{x}$ corresponding to the square modulus of 
the coefficients of the state in the computational basis 
($x \in \{ \ket{0\ldots 0}, \ket{0\ldots 01}, \ldots, \ket{1\ldots 1}\}$).
A quantum algorithm will be said to majorize
step by step this probability distribution if 
\begin{equation}
\label{maj6}
\vec p^{(m)} \prec \vec p^{(m+1)} \qquad \forall
m=1,\dots,M \ .
\end{equation}

In such a case, there will be a neat flow of probability directed
to the values of highest weight, in a way that the probability
distribution will be steeper and steeper as the algorithm  goes
ahead. This implies that the state is becoming less entangled along the 
computation. Notice that the majorization relations are stricter than
an inequality in the entanglement entropy, in such a way that
the reverse statement is not true.

In Ref.\ \cite{OLM04}, 
the step-by-step majorization 
was found in the known instances of fast and efficient algorithms, namely in the 
quantum Fourier transform, in Grover's algorithm, in the hidden affine 
function problem, in searching by quantum adiabatic evolution and in
deterministic quantum walks in  continuous time solving  a classically hard problem.
On the other hand, the optimal quantum algorithm for parity determination,
which does not provide any computational speed-up, does not 
show step-by-step majorization.

\begin{figure}
\begin{center}
\scalebox{0.4} 
{
\includegraphics*{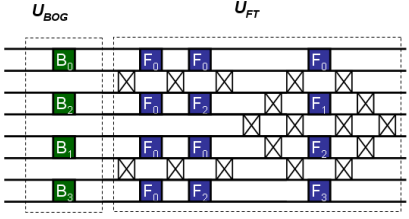}
}
\end{center}
\caption{
Structure of the quantum circuit performing the exact diagonalization of the
XY Hamiltonian for 8 sites. The circuit follows the structure
of a Bogoliubov transformation followed by a fast Fourier transform.
Three types of gates are involved: type-B (responsible for the
Bogoliubov transformation and depending on
the external magnetic field $\lambda$ and the anisotropy parameter
$\gamma$), type-fSWAP (depicted as crosses and necessary to implement the anti-commuting properties of fermions) and type-F
(gates associated to the fast Fourier transform). Some initial gates have been eliminated
since they only amount to some reordering of initial qubits [from \cite{VCL08}]. 
}
\label{fig:XX-circuit-n8}
\end{figure}

Recently, a new class of quantum algorithms have been presented.
Those are exact circuit that faithfully reproduce the dynamics
of strongly correlated many-body system.
In Ref.\ \cite{VCL08}, the underlying quantum circuit
that reproduces the physics of the  XY Hamiltonian for $N$ spins was obtained.
The philosophy inspiring that circuit was to follow the
steps of the analytical solution of that integrable model.
Looking at the architecture of the circuit in Fig.\ \ref{fig:XX-circuit-n8},
it is easy to realize that the entanglement between the two sets of
contiguous $N/2$ spins is transmitted through the $N/2$
SWAP gates.
Therefore, the maximum entanglement entropy between these two half's of the system
that this proposal may allow is $N/2$. 
This is because the maximum entanglement that can generate a quantum gate that acts on
two qubits is 1, that is, from a product state to a maximally entangled state (Bell basis).
Thus, the scaling law of the entanglement entropy that this proposal will
allow will be
\be
S(N/2) \le N/2 \, .
\ee
Notice that, as we have seen in the previous sections, 
the entanglement entropy of the ground state in the XY model scales only logarithmically.
The above circuit, then, can create much more entropy than what is present in the
ground state. Yet, we have also discussed the fact that time evolution
does create maximum entanglement. This, indeed, is what the above
circuit achieves.
This shows that the previous proposal is optimal since it carries the 
minimum possible number of gates such that maximum entanglement can
be created.

Let us also add a final example on exact quantum circuits.
In Ref.\ \cite{LPR09}, a quantum circuit that creates 
the Laughlin state (Eq.\ \ref{eq:Laughlin-def}) for an arbitrary
number of particles (qudits) $n$ in the case of filling fraction one is presented.
The way in which entanglement grows along
the circuit is also related to the amount of entanglement that each gate
of the circuit can generate. 
In the case of this Laughlin wave function, 
the depth of the circuit grows linearly with the
number of qudits, so  linear entanglement $S\sim n$ can be supported 
by the circuit. This is precisely the entanglement that the 
Laughlin wave function with filling fraction one requires as shown in Ref.\ \cite{ILO07}.

The exact circuits we have discussed (XY and Laughlin states) are
both able to create linear entanglement entropy. It is then impossible
they can be simulated classically in an efficient way.

\subsection{Adiabatic quantum computation}
The framework of adiabatic quantum computation (AQC) was introduced in Ref.\ \cite{FGGLLP01}.
The idea of AQC is the following:
\begin{enumerate}
\item A quantum register is initially prepared on the ground state
of a known initial Hamiltonian $H_0$.

\item The system is then made to evolve adiabatically
from this Hamiltonian to a new one $H_P$ whose ground state
codifies the solution to an {\sl e.g.} NP-complete problem
\be H(s(t))= (1-s(t)) H_0 + s(t) H_P \ . \ee

\item Slow evolution from $s(t=0)=0$  to $s(t=T)=1$ guarantees that
the system will not jump from the instantaneous ground state of the system to
the first excited state.
\end{enumerate}
Quantum adiabatic computation is proven
efficient provided that the minimum gap along the adiabatic
evolution is only polynomially small in the number of qubits.
If this was not the case, the adiabatic computation would  
require an exponentially large time as measure  in  terms
of the number of qubits in the register.

Thus, according to the previous arguments, at some point of
the adiabatic evolution of a hard quantum computation the system must be
highly entangled, in a similar way as it happened in the previous sections 
at the quantum phase transitions.
This makes us expect some sort of quantum phase transition for 
a concrete value $s_c$ of the Hamiltonian, 
point that would be characterized by a minimum energy gap.

In Ref.\ \cite{FGGS00}, adiabatic quantum computation is used 
to solve the NP-Complete Exact Cover problem that is a particular 
case of the 3-SAT problem. 
It is defined as follows: given the $n$ Boolean
variables $\{x_i\}_{i=1,\ldots n}$, $x_i = 0,1 \ \forall \ i$,
where $i$ is the bit index, we define a \emph{clause}
$C$ involving the three bits $i$, $j$ and $k$
 by the constraint $x_i + x_j + x_k = 1$. There are
only three assignments of the set of variables $\{x_i, x_j, x_k
\}$ that satisfy this equation, namely, $\{1,0,0\}$, $\{0,1,0\}$
and $\{0,0,1\}$.
An \emph{instance} of the Exact Cover problem is a collection of
clauses which involves different groups of three qubits. The
problem is to find a string of bits $\{x_1, x_2 \ldots , x_n \}$
which satisfies all the clauses.

This problem can be mapped to finding the ground state of a
Hamiltonian $H_P$ in the following way \cite{FGGS00}: given a clause $C$ define
the Hamiltonian associated to this clause as
\begin{align}
&H_C=\frac{1}{8}\left(
(1+\sigma_i^z)(1+\sigma_j^z)(1+\sigma_k^z) \right.\\ 
 &\left. + (1-\sigma_i^z)(1-\sigma_j^z)(1-\sigma_k^z)
 + (1-\sigma_i^z)(1-\sigma_j^z)(1+\sigma_k^z) \nonumber \right. \\   
 &\left. +(1-\sigma_i^z)(1+\sigma_j^z)(1-\sigma_k^z)
 + (1+\sigma_i^z)(1-\sigma_j^z)(1-\sigma_k^z) \right), \nonumber
\end{align}
where $\sigma^z |0\rangle = |0\rangle$, $\sigma^z
|1\rangle = -|1\rangle$.
The quantum states of the
computational basis that are eigenstates of $H_C$ with zero
eigenvalue (ground states) are the ones that correspond to the bit
string which satisfies $C$, whereas the rest of the computational
states are penalized with an energy equal to one.
The problem Hamiltonian is constructed as the sum of all the
Hamiltonians corresponding to all the clauses in the
instance,
\begin{equation}
H_P = \sum_{C \ \in \ {\rm instance}} H_C \ . 
\end{equation}
The ground state of this Hamiltonian corresponds to the quantum
state whose bit string satisfies \emph{all} the clauses.

It is known that Exact Cover is a NP-complete problem, 
so it cannot be solved in a polynomial number of steps in a classical computer \cite{CL70s}. 
This makes the Exact Cover problem, particularly interesting, 
since if we had an algorithm
to efficiently solve Exact Cover, we could also solve all problems
in the much larger NP family \cite{NPfamily}.

\begin{figure}
\centering
\includegraphics[angle=-90, width=0.49\textwidth]{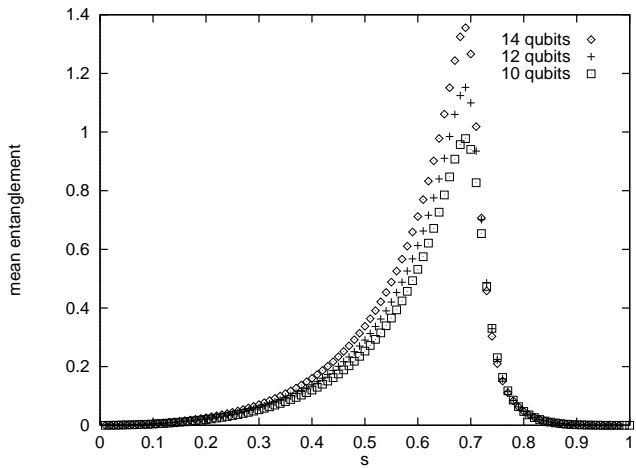}
\caption{Average over 300 instances of the entanglement entropy between two
blocks of size $n/2$  as a function
of the parameter $s$ controlling the adiabatic evolution. A
peak appears for $s_c \sim 0.7$. The plot also shows the increase of
the peak as the number of qubits grows $n=10,12,14$ 
[from \protect\cite{CC05}].} 
\label{fig:3sat}
\end{figure}

In Ref.\ \cite{LO05}, the evolution of the 
entanglement properties of the system are studied in order to see 
the expected sign of a quantum phase transition.
300 random instances for the Exact Cover are generated with only one
possible satisfying assignment for a small number of qubits.
This instances are produced by adding clauses at random until
there is exactly only one satisfying assignment. 
In order to apply adiabatic quantum computation the initial
Hamiltonian $H_0$ taken is a magnetic field in the $x$
direction
\begin{equation}
H_0 = \sum_{i = 1}^n \frac{d_i}{2}(1-\sigma_i^x) \ , \label{h0}
\end{equation}
where $d_i$ is the number of clauses in which qubit $i$ appears.
Then, for each instance, the ground state is computed 
for several values of $s$ 
of the Hamiltonian, $H(s)= (1-s) H_0 + s H_P$ and its 
corresponding entanglement entropy of half a chain.
The mean of the entanglement entropy over these 300 instances is performed
and plotted respect to the $s$ parameter for different 
sizes of the system in Fig.\ \ref{fig:3sat}. 
We can observe a peak of the entropy around the critical value $s_c\sim 0.7$.

We interpret this behaviour of the entanglement entropy as follows:
initially the system is in a product state and its entanglement is zero. 
Then, the evolution makes the system explore different solutions by means of
superposition states of them, that is, it becomes more and more entangled. 
Finally, the system throws away the bad solutions, the entanglement decreases, 
until the best solution is found and it rests in a product state again.
Roughly speaking, the power of the quantum computer respect to the classical one
underlies in the parallelism during the computation that the superposition principle allows.

Let us make some warning remarks. The numerical simulations performed for the
Exact Cover problem
cannot determine the complexity class for the quantum algorithm. It is
generally believed that quantum computers will no be able to handle
NP-complete problems. Yet, the simulation shows that the best this 
quantum algorithm can achieved still requires a huge amount
of entanglement in the register.

The divergence of the entanglement entropy that occurs at the critical point
$s_c$ have also been observed in Shor's
factoring algorithm in Ref.\ \cite{LatorreOrus04}, 
where entropy grows exponentially fast respect the number
of qudits. This, again,  makes this algorithm hard to simulate classically.

Notice that in the solution of other problems, the explosion of the entropy
could occur at $s_c=1$ in such a way that the entanglement entropy was monotonically 
increasing. This similar behaviour of the entropy to the quantum phase transition
is, therefore,  problem dependent.

Recently, there has been appeared a new quantum algorithm
for SAT problems that improve the previous results.
It consists of a hybrid procedure that alternates
non-adiabatic evolution with adiabatic steps \cite{GB08}.

\subsection{One way quantum computation}
The one-way quantum computation (or measurement based QC) 
is a method to perform quantum computation that consists of: 
({\it i}) first, an entangled resource state is prepared, 
and ({\it ii}) then single qubit measurements are performed on it. 
It is called "one-way" because the entanglement of the state, which is the resource of
the quantum computation, is destroyed by the measurements as the computation is being performed.
Although the output of each individual measurement is random, 
they are related in such a way that the computation always succeeds. 
The idea is that depending on the previous outcome, one chooses the basis for the next measurements.
This implies that the measurements cannot be performed at the same time.

This kind of computation was introduced in Ref.\ \cite{Oneway}
where there was shown that with an initial particular state, called cluster state,
any quantum computation could be simulated.
Later on, other useful states to perform one-way quantum computation were found 
\cite{UniversalShort, UniversalLong, Wires, Fundamentals,M}

The fact that the measurement based quantum computation is universal is
non-obvious, since a quantum computation is a unitary process, while
a measurement is a random process. The key point is that there are two 
kinds of qubits in the spin system: 
the cluster qubits which will be measured in the process of computation
and the logical qubits which constitute the quantum information
that is going to be processed.

Although, globally, entanglement is expected to decrease 
along the quantum computation due to
the single qubit measurements, in the set of logical qubits 
(the register that will be read out at the end of the computation),
the entanglement may increase. 
Notice that if the initial state fulfils an area law, the entanglement
is enough  that the register of the logical qubits is as entangled
as possible. That is, area law on a 2D state is just what is needed
to have linear maximal entanglement on a register defined on a line
in that state. Cluster states are just enough to handle the
expected entanglement in the register.
In this respect, it has been recently shown that 
most quantum states are too entangled to be useful in order to 
perform measured based quantum computation \cite{GFE09}.

\section{Conclusion: entanglement as the barrier for classical simulations}

Entanglement is the genuine quantum property that escapes
classical physics. The Hilbert space structure of a multi-partite
quantum system allows for
superpositions of exponentially many  elements of the basis.
Entropy of entanglement is a way to quantify the amount of
quantum correlation between parts of such a multi-partite system.
Entanglement entropy is, then, a genuine measure of the
global quantumness of the state.

It serves as a conclusion to recall the deep implications of entanglement
entropy in the possibility of producing faithful classical simulations
of quantum mechanics. In Ref.\ \cite{Vi03}, it was proven that efficient
simulations are possible for any system
where all its Schmidt decompositions in two arbitrary parts
would carry little entropy. Therefore, entanglement is at
the heart of the separation between efficient and non-efficient
simulations of quantum mechanics.

What is not fully understood
is what is the best strategy to classically account for quantum correlations.
Two general and clever ideas are available in the literature.
The first idea consists in exploiting the fact that typical interactions
are local. This suggests that entanglement
should be created sequentially in space from each local degree of freedom
to its nearest neighbours. Then, a one-dimensional state can be represented as
a matrix product state which captures such a principle \cite{PVWC07}. In higher
dimensions, states can be represented as Projected Entangled Pairs (see Refs.\ \cite{PEPS}).
The second idea to classically  represent quantum states
as efficiently as possible consists in reconstructing the correlations
in the system as a renormalization group tree. This goes under
the name of Multiscale Entanglement Renormalization Ansatz
(MERA) is a more
sophisticated representation which is specially suited
for critical systems. The accuracy of the approximations can
be quantify using the amount of entropy of entanglement
that the approximation can accommodate \cite{TOIL08}.

Multi-partite entanglement branches in many others subjects that
escape this short review. Very likely, much more work is still
needed to get a profound understanding of the role of entanglement
in highly structured quantum systems.

\begin{acknowledgements}
Financial support from QAP (EU), MICINN (Spain), 
FI program and Grup Consolidat  (Generalitat de Catalunya), 
and QOIT Consolider-Ingenio 2010 is acknowledged.
We also would like to thank the large number of colleagues
with whom we have collaborated and discussed  all along 
these recent years in the never-ending 
 effort to understand entanglement.
\end{acknowledgements}


\begin{thebibliography}{10}
\bibitem{NPfamily}
The NP problems are those whose solutions can be verified in a polynomial time.
\end{thebibliography}
\end{document}